\documentclass[prd,aps,preprint,showpacs]{revtex4}
\usepackage{amsfonts}
\usepackage{amsmath}
\usepackage{graphicx}
\def\hpsi{\hat{\Psi}}
\def\fn{f^{(n)}}
\def\psn{\psi^{(n)}}

\def\lsim{\mathrel{\raise.3ex\hbox{$<$\kern-.75em\lower1ex\hbox{$\sim$}}} }
\def\gsim{\mathrel{\raise.3ex\hbox{$>$\kern-.75em\lower1ex\hbox{$\sim$}}} }
\newcommand{\Z}{\mathbb{Z}_2}

\begin{document}
\preprint{TU-758}
\title{Hierarchical Mass Structure of Fermions in Warped Extra Dimension
\bigskip\bigskip}
\author{Sanghyeon Chang\footnote{schang@cskim.yonsei.ac.kr}}
\affiliation{Department of Physics and
IPAP, Yonsei University, Seoul 120-749, Korea}

\author{C. S. Kim\footnote{cskim@yonsei.ac.kr}}
\affiliation{Department of Physics and
IPAP, Yonsei University, Seoul 120-749, Korea}

\author{Masahiro Yamaguchi\footnote{yama@tuhep.phys.tohoku.ac.jp}}
\affiliation{Department of Physics, Tohoku University, Sendai 980-8578,
Japan
\bigskip\bigskip}

\begin{abstract}
\noindent The warped bulk standard model has been studied in the Randall-Sundrum
background on $S^1/\Z\times\Z'$ interval with the bulk gauge symmetry
$SU(3)\times SU(2)_L\times SU(2)_R\times U(1)_{B-L}$.
With the assumption of no large cancellation between the fermion flavor mixing
matrices, we present a simple analytic method to determine the bulk masses of standard model fermions
in the almost universal bulk Yukawa coupling model.
We also predict $U_{e3}$ element of  MNS matrix
to be near the experimental upper bound when the neutrino masses are of Dirac type.
\end{abstract}

\maketitle

\section{Introduction}

It is a fascinating idea that some of
the deep puzzles of particle physics may be attributed to the geometry
of extra space dimensions. The most discussed one is the gauge hierarchy
problem why the electroweak scale is much lower than the Planck
scale. An attractive hypothesis to explain this hierarchy was proposed
using large extra dimensions \cite{Arkani-Hamed:1998rs}.
Soon later an alternative interesting idea
was postulated by Randall and Sundrum \cite{Randall:1999ee}. In their
first model (RS1), the compact extra dimension has a size not much
larger than the Planck length, but with a warped metric. This warped
extra dimension is also interesting in the context of AdS/CFT
correspondence in string theories \cite{Maldacena:1997re}.
In fact, stringy realization of the
warped extra dimension was considered in compactifications with
non-vanishing fluxes of higher tensor fields. (See ref. \cite{Giddings:2001yu}
and references therein.)

Another puzzle which we would like to address here is the question of
fermion masses and  flavor mixing. An extra-dimensional explanation to this
puzzle owes to the configuration of the wave functions of the quarks and
leptons along the extra dimensions. In a field theory approach, the
smallness of the Yukawa coupling and thus the fermion mass is due to the
small overlap of the wave functions of the relevant fields in the extra
dimensions. The idea was proposed in flat TeV$^{-1}$ size extra dimension
\cite{Arkani-Hamed:1999dc}, which was utilized to construct realistic
models of the
Yukawa sector \cite{Kaplan:2001ga,Kakizaki:2001ue,Choi:2003di,Choi:2003ff}.

The geometrical approach to the Yukawa couplings  can also be applied to
the RS1 model. For this purpose, the standard model (SM) fields should
reside in the warped
bulk.  Though this was recognized to be possible
\cite{Davoudiasl:1999tf,Chang:1999nh}, the electroweak (EW) precision
test restricts the RS1 bulk SM strongly since the $t$ quark is much
heavier than other quarks and can give a significant amount of shift to
the weak gauge boson mass ratio $M_W/M_Z$ from  the SM prediction
due to $t$ quark and its
Kaluza-Klein (KK) mode mixing \cite{Hewett:2002fe}. Several attempts
were made to resolve this problem \cite{Hewett:2002fe, Kim:2002kk}.

Recently, Agashe {\it et al.} \cite{Agashe:2003zs} showed that the above problem of the too large
Peskin-Takeuchi $T$ parameter is due to the absence of a custodial $SU(2)$
symmetry in the bulk, as is suggested by the AdS/CFT correspondence, and
proposed a model which has the gauge symmetry $SU(3)_C \times SU(2)_L\times
SU(2)_R\times U(1)_{B-L}$.  This model may be related
with the warped Higgsless model which shows a possibility of EW symmetry
breaking
without a Higgs field in the RS1 SM \cite{Csaki:2003zu, Burdman:2003ya}.
It must be stressed that the Higgs field in the model we consider
must be confined on the brane
in order not to reincarnate the gauge hierarchy problem
\cite{Chang:1999nh}. Because of this peculiar property, the Higgs field
acts as a boundary condition (BC) for the bulk field equations.  If the
Higgs couples to different bulk fermions with (more or less) universal
strength at the boundary, the small masses and mixings of the SM
fermions can be induced by the suppression of the zero-mode wave
functions on the infra-red (IR) boundary \cite{Grossman:1999ra,
Gherghetta:2000qt}.

In this paper, we shall consider the fermion mass structure of quarks
and leptons (including neutrinos) in the framework of the warped bulk
fermions. Under the situation that a fundamental principle to dictate
the parameters in the 5D bulk theory is not known, it would be natural
to take the hypothesis that the 5D Yukawa couplings do not have any
particular textures. Thus we assume that the 5D Yukawa couplings are all
around unity in magnitude. Under this ``almost universal'' hypothesis,
the fermion mass structure is solely due to the configurations of the
(zero-mode) wave functions of the bulk fermions. In the case at hand,
they are controlled by bulk fermion masses. An attractive point of this
approach is that the whole bulk structure may be revealed in future
experiments to explore consequences of the Kaluza-Klein modes of the SM
particles.

The purpose of the paper is two fold: First we present a simple analytic
method which is useful to estimate the bulk fermion masses from known
experimental data under the assumption of the almost universal 5D Yukawa
couplings. Despite the fact that there are already many (numerical)
analyzes on the fermion masses in the warped extra dimension in the
literature, we believe that it is still worth presenting our analytic
results because of simplicity and accessibility. The hierarchical structure
in the quark mass matrices makes our analysis very robust. For the lepton
sector, although it may suffer from some pollution of numerical
coefficients because of its somewhat less hierarchical pattern of the
masses and  mixings, it is still possible to determine generic structure.
The second purpose of the paper is to show that the $U_{e3}$ entry of
the Maki-Nakagawa-Sakata (MNS) mixing matrix in the neutrino sector is
typically close to the present experimental upper bound when the neutrino
masses are of Dirac type.

\section{The standard model in the RS1 bulk}

The basic framework of our study is a simple system where one flavor of
fermion resides in the bulk of the RS1. We extend it to the three flavor
system later in this paper.
The RS1 metric is given by
\begin{equation}
ds^2= e^{-2\sigma(y)}(dt^2-dx^2) - dy^2,
\end{equation}
where $y$ represents the warped coordinate for extra-dimension
and $\sigma(y) = k|y|$. The 5th dimension is bounded in the interval
$(0,L)$. The gravity is confined at $y=0$ boundary known as Planck (UV) brane,
whereas our world is confined on the other end ($y=L$), which is called
the TeV (IR) brane.
$y$ coordinate can be converted to a conformally flat coordinate,
$z\equiv e^\sigma/k$, where the metric becomes
\begin{equation}
ds^2= \frac{1}{(kz)^2}(dt^2-dx^2 - dz^2).
\end{equation}
The interval becomes $1/k < z <1/T$, where $T=k e^{-kL}\sim {\cal O}(1) $ TeV .

The 5D fermion action becomes
\begin{eqnarray}
S_{\rm fermion}=\int d^4x dy \left[ \overline{\hat\Psi} e^\sigma
i\gamma^\mu\partial_\mu \hat{\Psi} -\frac{1}{2}
\overline{\hat\Psi} \gamma_5\partial_y \hat{\Psi}
+\frac{1}{2}(\partial_y \overline{\hat\Psi})\gamma_5 \hat\Psi
 + m_D \overline{\hat\Psi}\hat\Psi
 + m_M \overline{\hat\Psi}\hat\Psi^c
 \right]\ ,
\end{eqnarray}
where $\hat{\Psi}\equiv e^{-2\sigma}\Psi .$
The bulk
Majorana mass $m_M$ is non-zero only if the fermion is neutral.
This Lagrangian has the five-dimensional $\Z\times\Z'$ parity
on each boundary (brane),
\begin{equation}
\gamma_5 \Psi (x,-y)\,={}\pm\Psi (x,y) \, ,\hspace{6mm}
\gamma_5 \Psi (x,L-y)\,={}\pm\Psi (x,L + y) \ .
\end{equation}
$\Z$ and $\Z'$ represent UV and IR parity of bulk field and written in the
form of (UV, IR).
The bulk Dirac mass is defined by $m_D=\sigma'= kc\ {\rm sign}(y)$.
The bulk fermion can be divided into two chiral components,
$\hpsi=\hpsi_L+\hpsi_R$, for $ \gamma_5\hpsi_L= -\hpsi_L \, ,
\gamma_5\hpsi_R= \hpsi_R$.
Each chiral field can be expanded to the KK modes
\begin{equation}
\hat{\Psi}(x,y)_{L(R)} = \sqrt{k}\sum_n \psi_{L(R)}^{(n)}(x) f_{L(R)}^{(n)}(y) .\
\end{equation}
After the mode expansion, the 4D effective action for KK modes becomes
\begin{equation}
S_{\rm eff}
 = \int d^4 x \sum_n\left[\overline{\psi}^{(n)}_L i \gamma^\mu \partial_\mu
 \psi^{(n)}_L + \overline{\psi}^{(n)}_R i\gamma^{\mu} \partial_\mu
\psi^{(n)}_R - m^{(n)}(\overline{\psi}^{(n)}_L \psi^{(n)}_R +
\overline{\psi}^{(n)}_R \psi^{(n)}_L)
     \right] \ ,
\label{eff0}
\end{equation}
where $m^{(n)}$ is a mass of nth KK excited mode.
To generate the action (\ref{eff0}), KK mode functions should
satisfy the mode equations in $z$ coordinate,
\begin{eqnarray}
\left(\partial_z \pm \frac{c}{z}\right) f_{L/R}^{(n)} = \mp m^{(n)}
f_{R/L}^{(n)}~ .
\end{eqnarray}
%
A generic 5D bulk fermion can have
four different forms according to the $\Z \times \Z'$ parity,
\begin{eqnarray}
\label{eq:Psi1} \hpsi_i(x,y) & = & \sqrt{k}\sum_n [\psn_{iL}(x)
\fn_{iL}(y)
+ \psn_{iR}(x) \fn_{iR}(y) ] \, .
\end{eqnarray}
The indices $i=1,2$ represent the parallel conditions, where
$f_{iL}$ has $(\pm\pm)$ parity  and $f_{iR}$ has $(\mp\mp)$, and
$i=3,4$ represent the crossed conditions, where $(\pm\mp)$ for
$f_{iL}$ and $(\mp\pm)$ for $f_{iR}$, respectively. Each mode
function except the zero modes can be written in the series of
Bessel functions. For more details, see Ref. \cite{Chang:2005vj}.

The Higgs field $\phi(x)$ is confined on the IR boundary,
\begin{eqnarray}
S
=-\int d^4 x  dy  \frac{\lambda_{5}}{T}
H(x)\left(\overline{\hpsi}_{1}(x,y)
\hpsi_{2}(x,y)+\overline{\hpsi}_{2}(x,y)
\hpsi_{1}(x,y)\right)\delta(y-L)~,
\end{eqnarray}
where $H=e^{- k L}\phi(x)$ is canonically normalized Higgs
scalar and $\lambda_5$ is the Yukawa coupling. When the Higgs field
get a vacuum expectation value $\langle H \rangle = v_W$,
the surviving zero modes give the SM fermion mass term,
\begin{equation}
m_f=\left.\frac{\lambda_5 v_W k}{T}f_{1L}^{(0)}f_{2R}^{(0)}\bar\psi^{(0)}_{1L}
\psi^{(0)}_{2R} \right|_{z=1/T},
\label{chiralmass}
\end{equation}
where the zero mode functions are
\begin{eqnarray}
f_{1L}^{(0)} = \frac{(kz)^{-c_1}}{N_1^{(0)}},\ \  ~~~
f_{2R}^{(0)} = \frac{(kz)^{c_2}}{N_2^{(0)}},
\end{eqnarray}
and the normalization becomes
\begin{eqnarray}
N^{(0)}_1=\sqrt{\frac{ 1- \epsilon^{2{c}_1-1}}{2{c}_1-1}} , \ \  ~~~
N^{(0)}_2=\sqrt{\frac{  \epsilon^{-2{c}_2-1}-1}{2{c}_2+1}} ,
\end{eqnarray}
with $\epsilon= T/k=e^{-kL}$. We will drop the  indices
1 and 2 from  this point to avoid the confusion with
family  indices.

The SM requires that two $SU(2)_L$ singlet right-handed fermions
should exist for a
corresponding left-handed doublet.  To match the particle content,
we set  $(Q_i,U_i,D_i)$ and $(L_i,N_i,E_i)$ as bulk fields
where $i=1,2,3$ represent 3 generations.
$Q$ and $L$ include $SU(2)_L$ quark and lepton doublets.
$U,D,E,N$ include the SM fields $(u,d,e)_R$ and a right-handed neutrino $N_R$,
respectively.

If we expand the model to 3 generations, the mass term is written in  $3\times 3$
matrix,
\begin{equation}
M^f_{ij}/v_W =    y^f_{ij} =\lambda^f_{5ij} F_L(c_i) \times F_R(c_j) ,
\end{equation}
where  $y^f_{ij}$ is a 4D effective Yukawa coupling of fermion $f$ and
$\lambda^f_{5ij}$ is a 5D boundary Yukawa coupling, and
\begin{equation}
F_{L}(c_i) = \epsilon ^{c_i-1/2} \sqrt{\frac{ 2c_i -1}{1- \epsilon^{2c_i -1}}}
,\ \  ~~~
F_{R}(c_i) = \epsilon ^{-c_i-1/2} \sqrt{\frac{ 2c_i +1}{\epsilon^{-2c_i
-1}-1}},
\label{para1}
\end{equation}
where $c_i$ represents each mass of $(Q_i,U_i,D_i)$
and $(L_i,N_i,E_i)$.
$F_{L(R)}(c_i)=1$ when bulk fermion mass is zero $c_i=0$. If we increase
$(-)c_i$,
$F_{L(R)}(c_i)$ decrease slowly until $(-)c_i=1/2$. For $(-)c_i>1/2$,
it decrease fast in power of  $\epsilon ^{(-)c_i}$;
\begin{eqnarray}
F_{L(R)}(c_i) &\simeq  &  \epsilon ^{(-)c_i-1/2} \sqrt{(-)2 c_i -1}
\hspace{1cm} \mbox{ for } (-)c_i-1/2 \gg 1/kL
\nonumber \\
 & \simeq &  (kL)^{-1/2}\hspace{2.8cm}  ~~~ \mbox{ for } |(-)c_i -
 \frac{1}{2}| \ll 1/kL
\nonumber \\
 & \simeq &  \sqrt{ 1-(-)2c_i} \hspace{2.1cm} ~~~ \mbox{ for } (-)c_i -1/2
\ll -1/kL .
\end{eqnarray}
The mass difference of bulk fermion gives the natural mass
hierarchy between different SM fermions.

\section{Fermion Masses and Mixings}

The bulk SM conflicts with the electro-weak precision test without some
symmetry \cite{Hewett:2002fe,Kim:2002kk}. The $SU(2)_L\times SU(2)_R\times U(1)_{B-L}$ bulk SM is a favorable
candidate  because  its custodial isospin prevents the extra-contribution
from the KK fermion modes to the gauge boson self-energy \cite{Agashe:2003zs}.
Also, this model draws interests due to the connection with the Higgsless
model of electro-weak symmetry breaking \cite{Csaki:2003zu,Burdman:2003ya}.

All SM fields except the Higgs field reside in the bulk \cite{Chang:1999nh}.
There are some fields which have no SM counter part, $e.g.$ $SU(2)_R$ charged
gauge bosons. The ``crossed" BC $(\pm\mp)$ assigned to these  fields eliminates
their zero modes, thus we will not see any light additional field.
Among the bulk fermions we defined in previous section,
$Q$ and $L$ fields are consisted with $(\pm\pm)$ fields only,
while $SU(2)_R$ doublet  contains one component with $(\pm\mp)$ parity,
because their charged current should conserve $\Z\times\Z'$ parity.
Thus, for $U,N, D,E$ fields, only one  component of the doublet
can have a zero mode.
If the SM is induced from this model, there should be at least one bulk
$SU(2)_L$ doublet and two $SU(2)_R$ doublet fermions for each family.

To establish a simple but realistic model for 4D fermion
masses,
 we choose the bulk mass matrices are real and diagonal.
Also, for simplicity
we ignore CP phase in the Yukawa couplings. Inclusion of the CP
phase is straightforward.
In this paper, we use a (almost) universal Yukawa coupling model that
the Higgs scalar couples to all fermions with (almost) universal
strength.
In this model, the fermion mass hierarchy is  generated only by
the bulk fermion mass structures.
For the case that the universality is exact,
$ 3\times 3$ matrix $M_{ij}=v_W F_L(c_{Qi}) F_R(c_{Aj})$ has
only one  non-zero eigenvalue.
This approach is similar to the fermion mass hierarchy generation
method by Froggatt and Nielsen which was used in anomalous $U(1)$ model \cite{Froggatt:1978nt,Elwood:1997pc,
Elwood:1998kf}, and also in various bulk SM models
\cite{Agashe:2004ay,Agashe:2004cp,Huber:2003tu,Choi:2003ff,Choi:2003di,Moreau:2005kz}.

\subsection{Quark Masses and Mixings}

The bulk  fields $Q_i$, $U_i$ and $D_i$
with bulk mass parameters $c_{Qi}$, $c_{Ui}$, $c_{Di}$,
contain the zero modes which can be interpreted as the SM fermions.
If we take all parameters to be real, the mass matrices can be diagonalized by
bi-orthogonal transformation,
\begin{equation}
U^T_{qL} M_q U_{qR}=M_q^{\rm diag} ~~~~~ \mbox{ for }
q=u,d.  \label{bioth}
\end{equation}
The CKM matrix is defined as $K= U_{uL}^T U_{dL}$.
With simplified Wolfenstein parametrization
for  $\lambda \simeq 0.22$, the CKM matrix $K$ can be written
\begin{equation}
K\simeq
\left(\begin{array}{ccc} 1 &\lambda & \lambda^3 \\
\lambda & 1 & \lambda^2 \\ \lambda^3 & \lambda^2 & 1
\end{array} \right), \label{CKM}
\end{equation}
where the numerical coefficient of each entry is of order unity. A
natural choice for $U_{u_L}$ and $U_{d_L}$ in this case is  of the
similar form as (\ref{CKM}):
\begin{equation}
  U_{u_L} \simeq U_{d_L} \simeq
\left(\begin{array}{ccc} 1 &\lambda & \lambda^3 \\
\lambda & 1 & \lambda^2 \\ \lambda^3 & \lambda^2 & 1
\end{array} \right). \label{mixing-quark}
\end{equation}
Here any number greater than $\lambda^{0.5}$ is replaced as  unity.
The above choice of mixing is reasonable since the $u_L$ and $d_L$ has the
same bulk mass.
The fermion masses can also be expressed in terms of $\lambda$,
\begin{eqnarray}
M_u^{\rm diag} = diag(m_u, m_c, m_t) &\simeq& v_{W}\  diag(\lambda^8, \lambda^{3.5}, 1),
\nonumber\\
M_d^{\rm diag} = diag(m_d, m_s, m_b) &\simeq& v_{W}\  diag(\lambda^7, \lambda^5, \lambda^{2.5} ).
\label{quarkmass}
\end{eqnarray}
If we consider the (almost) universal coupling, the quark mass matrices become
\begin{equation}
(M_a)_{ij}\simeq v_W  F_L(c_{Qi}) F_R(c_{Aj}), \label{ansatz}
\end{equation}
where $a=u,d$ and $A=U,D$.
It follows from the above that
\begin{eqnarray}
(M_aM_a^T)_{ij}=  (U_{aL} (M_a^{D})^2 U_{aL}^T)_{ij}
\simeq
v_W^2 F_L(c_{Qi}) F_L(c_{Qj}) (\sum_k F_R(c_{Ak})^2)
.
\end{eqnarray}

Let us now determine the mass parameters $c$'s.
For $u$ quark, we find
\begin{eqnarray}
M_u M_u^T
\simeq (v_W C)^2\left( F_L(c_{Qi}) F_L(c_{Qj}) \right) \simeq v_W^2
\left(\begin{array}{ccc} \lambda^6 &\lambda^5 & \lambda^3 \\
\lambda^5 & \lambda^4 & \lambda^2 \\ \lambda^3 & \lambda^2 & 1
\end{array} \right),
\end{eqnarray}
where the last equality is obtained by substituting
Eqs.~(\ref{mixing-quark}) and (\ref{quarkmass}) into (\ref{bioth}).
This leads
\begin{eqnarray}
F_L(c_{Q1})  \simeq  C^{-1}\lambda^3, ~~~~
 F_L(c_{Q2})  \simeq C^{-1} \lambda^2, ~~~~
 F_L(c_{Q3})  \simeq  C^{-1},
\end{eqnarray}
where $C\simeq F_R(c_{U3})$. Notice that this procedure works because of
the hierarchical mass structure $m_t \gg m_c, m_u$. This observation is
crucial when discussing the neutrino masses. We will come back this
point shortly.

If $c_{U3}$ is too large, the mass of down sector quark from $SU(2)_R$
doublet $U_3$ becomes too small, giving too much contribution to
Peskin-Takeuchi $T$ parameter. It should be restricted,  $F_R(c_{U3})
\lsim 1.2$.  Also the constraint from $Z\rightarrow b\bar{b}$ gives the
allowed range $F_L(c_{Q3}) \lsim 0.7$ \cite{Agashe:2003zs}.  Since
$m_t/v_W \simeq F_L(c_{Q3}) F_R(c_{U3})\simeq 1$, for the range of our
interest, 2 TeV $<T<$ 8 TeV and with the standard choice of curvature
scale $k\simeq M_{pl}$, the bulk top masses are almost fixed around the
values  $c_{U3}\simeq 0.2$ and $c_{Q3} \simeq 0.3$.  Then from
(\ref{para1}), we find
\begin{eqnarray}
c_{Q1} \simeq 0.61,\  ~~~~
c_{Q2}  \simeq 0.56 ,\  ~~~~
c_{Q3} \simeq 0.3 .
\end{eqnarray}
If we assume that off-diagonal term in $U_{qR}$ is small enough, with
Eq. (\ref{bioth}) and (\ref{quarkmass}), the quark mass matrices can be written
in the following form:
\begin{eqnarray}
M_u
\simeq v_W
\left( \begin{array}{ccc}\lambda^8 &\lambda^{4.5} & \lambda^3 \\
                         \lambda^9  & \lambda^{3.5} & \lambda^2 \\
                         \lambda^{11} & \lambda^{5.5}  & 1
\end{array}
\right) U_{u_R}^T
\simeq v_W
\left(\begin{array}{ccc} \lambda^8 &\lambda^{4.5} & \lambda^3 \\
 & \lambda^{3.5} & \lambda^2 \\  &  & 1
\end{array} \right),\
\label{uquark}
\end{eqnarray}
for the $U$ fields and,
\begin{eqnarray}
M_d
\simeq v_W
\left(\begin{array}{ccc} \lambda^7 &\lambda^6 & \lambda^{5.5} \\
                         \lambda^8 & \lambda^5 & \lambda^{4.5} \\
                         \lambda^{10}& \lambda^7  & \lambda^{2.5}
\end{array} \right) U_{d_R}^T
\simeq v_W
\left(\begin{array}{ccc} \lambda^7 &\lambda^6 & \lambda^{5.5} \\
 & \lambda^5 & \lambda^{4.5} \\  &  & \lambda^{2.5}
\end{array} \right),
\label{dquark}
\end{eqnarray}
for the $D$ fields.
The lower left components
depend on the details of the mixing matrices and
are redundant for the mass determination.
Our hypothesis of the almost universal 5D Yukawa couplings implies that
both of the mass matrices given above are expressed as
Eq.  (\ref{ansatz}), which can be achieved if one chooses
\begin{eqnarray}
c_{U1} \simeq -0.70 ,\  ~~~~
c_{U2} \simeq -0.52 ,\  ~~~~
c_{U3} \simeq 0.2 ,
\end{eqnarray}
\begin{eqnarray}
c_{D1} \simeq -0.65 ,\  ~~~~
c_{D2} \simeq -0.60 ,\  ~~~~
c_{D3} \simeq -0.57 .
\end{eqnarray}
There can be small modification for a different choice of  initial
parameter range.
The bulk masses we obtained above are approximately in agreement with
the previous calculations
\cite{ Agashe:2004cp,Huber:2003tu}.

\subsection{Lepton Masses and Mixings}

We now consider the mass matrices for charged leptons and neutrinos.
It is required to use a more cautious analysis to the lepton sector,
because the hierarchy between lepton masses and mixings is weaker than
that of quarks.

An advantage of the extra-dimensional explanation for the fermion masses
 is that the small masses can easily be generated as a consequence of
 the separation of the wave functions. When the fermions are in the
 warped extra dimension, the zero-mode wave functions have exponential
 form so that this suppression mechanism is very effective. Thus the
 Dirac masses of the neutrinos can be very small, allowing us to discuss
 the case where the light neutrinos are Dirac ones.

Motivated by the aforementioned argument, let us first consider the
Dirac neutrino case.  We assume $SU(2)_L$ doublet bulk leptons $L_i$
with bulk mass $c_{Li}$ and $SU(2)_R$ doublets $E_i$ and $N_i$ with
masses $c_{Ni}$ and $c_{Ei}$. Each of them contains the zero mode
$l_{iL}$, $e_{iR}$ and $\nu_{iR}$, respectively.  If we consider that the
SM neutrinos are Dirac particles, the MNS mixing matrix for neutrino is
equivalent to the CKM matrix, $U_{MNS}= U_e^\dagger U_\nu$, where
\begin{equation}
M_\nu^\dagger M_\nu  = U_\nu (M_\nu^{\rm diag})^2U_\nu^\dagger ,\ \  ~~~
M_e^\dagger  M_e = U_e (M_e^{\rm diag})^2 U_e^\dagger .
\end{equation}
for Dirac neutrino mass.
With the same approximation as the quark case,
the MNS matrix can be approximated as
    \begin{eqnarray}
   | U_{\rm MNS}| \sim \left(\begin{array}{ccc}
 1 & 1 & \lambda^m \\
1 & 1& 1 \\
1 & 1& 1
    \end{array}\right),
    \label{mixing}
    \end{eqnarray}
where the experimental constraint on $U_{e3}$ gives $m>1.3$.

Though the individual neutrino masses are not yet measured,
the  mass differences between them are determined by the neutrino
oscillation data,
\begin{eqnarray}
\Delta m_{sol}^2 = m^2_2 - m^2_1\simeq 7.5\times
10^{-5}~\mbox{eV}^2, \  ~~~~
  \Delta m_{atm}^2= |m^2_3 - m^2_2|\simeq  2.5\times
10^{-3}~\mbox{eV}^2.
\end{eqnarray}
The WMAP result  suggests
that any of neutrino mass should be $m_i < 1.0$ eV.  With all known data,
there exist three possible cases: (1) almost degenerate neutrinos, (2) the
normal hierarchy (NH), (3) the inverse hierarchy (IH).  If the neutrino
masses are almost degenerate $m_i\lesssim 1$ eV, then with the maximal
mixing of the MNS matrix, we expect that all the left-handed mode
functions have almost the same configurations.  Also the right-handed
neutrinos should have the same pattern, while the right-handed charged
lepton should have the hierarchical form.  This may be possible.
However, the structure of the MNS matrix as well as the mass differences
would be a consequence of some numerology. We will not discuss this case
furthermore.

For the NH case, as $\nu_1$ is very
light or even massless, the other neutrino masses are fixed as
  \begin{eqnarray}
        m_1=0 ,\  ~~~
        m_2=\sqrt{\Delta m_{sol}^2} ,\  ~~~
        m_3=\sqrt{\Delta m_{sol}^2+\Delta m_{atm}^2}.
    \end{eqnarray}
For the IH, $\nu_3$ is very light so that
  \begin{eqnarray}
            m_1=\sqrt{\Delta m_{atm}^2 -
                \Delta m_{sol}^2} ,\  ~~~
            m_2=\sqrt{\Delta m_{atm}^2},\  ~~~
            m_3=0 .
    \end{eqnarray}
If we allow the random cancellation during the diagonalization of mass matrix,
there can be too many possibilities. On the other hand, if we follow the first
assumption of no-cancellation strictly, the mass matrix should have either of
the following two forms
\begin{equation}
M_\nu^T M_\nu\propto
\left(\begin{array}{ccc} \lambda^{2n} &\lambda^n &
\lambda^n \\
\lambda^n & 1 & 1 \\ \lambda^n & 1 & 1
\end{array} \right)~~ \mbox{(NH)~~~~ or~~~~ }
  \left(\begin{array}{ccc} \lambda^{2k} &1 &
1 \\
1& \lambda^{2l} & \lambda^{2l}   \\
1& \lambda^{2l} & \lambda^{2l}
\end{array} \right)~~ \mbox{(IH) },
 \label{numass}
\end{equation}
where $k$, $l$ and $n$ are some positive numbers.
The derivation of the above can be found in
Refs.~\cite{Altarelli:2002hx,Altarelli:2002sg}.
In short, we utilize the fact that $U_{\rm MNS}$ is almost tri/bi-maximal and
the neutrino masses are close to
 (0,0,1) for NH and  (1,-1,0) or (1,1,0)  for IH.
We can derive (\ref{numass}) by adding a small perturbation to the solutions of
the approximation. For the IH case, $k\sim l\gsim 1$ is favored to avoid too large
$U_{e3}$.

It is clear that the IH is not consistent with
our almost universal Yukawa coupling approach, where
the lepton mass matrices should be written as
\begin{equation}
(M_\nu)_{ij} \simeq v_W  F_L(c_{Li}) F_R(c_{Nj}), \ \  ~~~~
(M_e)_{ij} \simeq  v_W F_L(c_{Li}) F_R(c_{Ej}).
\end{equation}
Therefore we consider only the NH case, where the lepton masses can be written as,
\begin{eqnarray}
M_\nu^{\rm diag} = diag(m_1, m_2, m_3) &=& v_W\  diag(<\lambda^{20.5}, \lambda^{20.5}, \lambda^{19}),
\nonumber\\
M_e^{\rm diag} = diag(m_e, m_\mu, m_\tau) &=& v_W\  diag(\lambda^{8.5}, \lambda^5, \lambda^3 ).
\label{1.5}
\end{eqnarray}
Our hypothesis implies as in the quark sector
\begin{eqnarray}
(M_a^T M_a)_{ij} &\simeq& (U_a (M_a^D)^2 U_a^T)_{ij}
\simeq v_W^2 F_L(c_{Li}) F_L(c_{Lj}) \sum_k F_R(c_{Ak})^2
\label{lepton2}
\end{eqnarray}
with $a=\{\nu, e\}$ and $A= \{N, E\}$, which can accord with the NH
neutrino masses. With Eq.~(\ref{numass}), one finds
\begin{equation}
M_e^T M_e
\simeq  v_W^2\lambda^{6} \left(\begin{array}{ccc} \lambda^{2n} &\lambda^n &
\lambda^n \\
\lambda^n & 1 & 1 \\ \lambda^n & 1 & 1
\end{array} \right),\ \  ~~~~
 M_\nu^T M_\nu\simeq \lambda^{32} M_e^T M_e~.\label{Un}
\end{equation}
The  bulk mass terms of $SU(2)_L$ doublets are
\begin{eqnarray}
 F_L(c_{L1} )\simeq  C_L^{-1} \lambda^{3+n},\  ~~~~
 F_L(c_{L2} ) \simeq  C_L^{-1} \lambda^{3} ,\  ~~~~
 F_L(c_{L3} ) \simeq  C_L^{-1} \lambda^{3},
\end{eqnarray}
where  $C_L\sim  F_R(c_{E3})$.

Unlike the quark case, we cannot simply set the mixing matrices $U_e\simeq
U_{\nu}$.
The maximal mixing between 2 and 3 flavors together with
(\ref{Un}) suggests the following left-handed mixing matrices
\begin{equation}
U_f\simeq \left(\begin{array}{ccc}
1 & \lambda^{a_f} & \lambda^n \\
\lambda^{b_f}  & 1& 1 \\
\lambda^{c_f}  & 1& 1
    \end{array}\right), \label{Uf}
\end{equation}
with $f=e,\nu$.
Writing
\begin{equation}
    (M_a^T M_a)_{ij}=m_{a3}^2 U_{ai3}U_{aj3}
   +m_{a2}^2 U_{ai2}U_{aj2}+m_{a1}^2 U_{ai1}U_{aj1},
\end{equation}
with $m_{ai}$ being the $i$-th mass eigenvalue of species $a$, one finds
that the first term in the right-handed side should dominate over the
rest to reproduce (\ref{Un}). This requires that the mass eigenvalues are
more hierarchical than the mixings. In fact, one finds
\begin{equation}
  1.5+a_{\nu} \gtrsim n,  2+a_{e} \gtrsim n. \label{hierarchy}
\end{equation}
Next we consider the MNS matrix.
The  MNS matrix  $U_{\rm MNS}=U^T_e U_\nu$ can be evaluated by using
(\ref{Uf}).
Then
Eq. (\ref{mixing}) implies
\begin{eqnarray}
         \lambda^{a_{\nu}}\sim \lambda^{b_{\nu}}+\lambda^{c_{\nu}} \sim
      1,\  ~~~~
         \lambda^{b_{e}}+\lambda^{c_{e}} \lesssim \lambda^{m},\  ~~~~
         \lambda^{n}\lesssim \lambda^{m} \label{UeUnu}.
\end{eqnarray}
Eqs. (\ref{hierarchy}) and (\ref{UeUnu}) restrict the allowed values  of
$n$ and $m$ in a narrow range
\begin{equation}
  1.3 \lesssim m \lesssim n \lesssim 1.5.
\end{equation}
Thus, as a representative value, we expect
\begin{equation}
  U_{e3} \simeq \lambda^m \simeq 0.10-0.14,
\end{equation}
provided that the 5D Yukawa couplings are almost universal and no
accidental cancellation takes place in the determination of the mass
structure. This value is close to the present upper bound and should be
explored by near future experiments. This observation  may be one of the most
important consequences of the present paper. For $m\simeq 1.5$, it is
interesting to rewrite the above as follows:
\begin{equation}
   U_{e3} \simeq \sqrt{\frac{\Delta m_{sol}^2}{\Delta m_{atm}^2}}.
\label{interesting-relation}
\end{equation}
We should note that the numerical coefficient in front cannot be
determined in our framework. The actual value would depend on the range
of the 5D Yukawa couplings. We should also note that the interesting
relation (\ref{interesting-relation}) may be polluted by possible
cancellation among various contributions because of the less
hierarchical structure of the neutrino masses and  mixings.

The charged lepton bulk mass can be obtained with the similar method as
the quark case. The relations,
\begin{eqnarray}
F_L(c_{L1}) F_R(c_{E1}) \simeq   \lambda^{8.5},\  ~~~~
 F_L(c_{L2}) F_R(c_{E2})  \simeq  \lambda^5,\  ~~~~
 F_L(c_{L3}) F_R(c_{E3})  \simeq  \lambda^3,
\end{eqnarray}
hold approximately if the right-handed lepton mixing is chosen to be
 small enough. However, there is  a wider range of solution space in above
equation than that of quarks.
The  lepton flavor violation
limit from $\mu \rightarrow 3e$ and other
experimental data  restrict that $c_{L3}$ cannot be smaller
than $0.5$ \cite{Huber:2003tu}.
The lower bound on the bulk lepton masses are,
\begin{eqnarray}
c_{L1}  \simeq 0.59,\  ~~~~~
&&c_{L2}   \simeq 0.5 ,\  ~~~~~
c_{L3}  \simeq 0.5 ,\nonumber\\
c_{E1} \simeq -0.74,\  ~~~~~
&&c_{E2}  \simeq - 0.65 ,\  ~~~~~
c_{E3} \simeq  -0.55.
\end{eqnarray}
On the other hand, for $c_{E3} \simeq 0$, one finds
\begin{eqnarray}
c_{L1} \simeq 0.68,\  ~~~~~
&&c_{L2}  \simeq 0.61 ,\  ~~~~~
c_{L3} \simeq 0.61 ,\nonumber\\
c_{E1} \simeq -0.65,\  ~~~~~
&&c_{E2}  \simeq - 0.55 ,\  ~~~~~
c_{E3} \simeq 0 .
\end{eqnarray}
There is no strict upper bound on the bulk masses, but if $c_{E3}$
is much larger than 0.5, the  first KK neutrino in $SU(2)_R$
doublet $E$, which has $(+-)$ BC, becomes too light. For instance,
$c_{E3}\simeq 0.7$ leads $m_N^{(1)}\sim 1$ GeV and if $c_{E3}$
approaches to the value $1$, the mass become lower than MeV and
can be considered as a sterile neutrino. This type of neutrino KK
mode may conflict with experimental or cosmological data
\cite{Chang:2005vj,Agashe:2004bm}.

Even though we fix $\nu_2$ and $\nu_3$ masses in the NH case,
there is no data which determines the lightest neutrino mass.
In other  words,  two right-handed neutrinos are just enough to explain all
the existing experimental data. Using the similar method, we can determine
the two bulk neutrino masses with the relations,
\begin{eqnarray}
 F_L(c_{L2}) F_R(c_{N2})  \simeq   \lambda^{17.5} ,\  ~~~~
 F_L(c_{L3}) F_R(c_{N3} ) \simeq   \lambda^{16}.
\end{eqnarray}
In the valid range where $c_{L3} >0.5$, the lower bound becomes,
\begin{eqnarray}
c_{N2}  \simeq -1.2 ,\  ~~~~
c_{N3} \simeq -1.1 .
\end{eqnarray}
This value does not vary much in the range $0.5 <c_{L3}<0.6$. Note that
$c_{E3}$ is the most sensitive parameter and might be the easiest one
to test at the near future high energy experiments.

Finally, we examine the case where
the Majorana mass $m_M=\lambda_L \delta(z-1/k)$ is present at the UV brane.
It is known that the neutrinos can acquire a small Majorana mass via
bulk seesaw mechanism even
for a small $\lambda_L >10^{-11}$ \cite{Huber:2003sf},
\begin{eqnarray}
 (M_\nu)_{ij} \simeq  v_{W}^2 \sum_{kl} h^\nu_{ik} h_{jl}^\nu F_L(c_{Li})
F_R(c_{Nk})
M_{Rkl}^{-1} F_L(c_{Lj})
 F_R(c_{Nl}).
\end{eqnarray}
The Majorana mass matrix  can be written as
\begin{eqnarray}
M_{Rij} \simeq \frac{\lambda_{Lij}}{2} k F_R(c_{Ni})
F_R(c_{Nj})\epsilon^{-c_{Ni}-c_{Nj}+1}  .
\end{eqnarray}
The assumption of the coupling universality for both boundaries,
$\lambda_{Lij}\sim \lambda_L$ leads,
\begin{eqnarray}
 (M_\nu)_{ij} \simeq \frac{F_L(c_{Li}) F_L(c_{Lj}) v_{W}^2
}{\lambda_L T}
  \epsilon^{ 2 c_{N1} },
\end{eqnarray}
where $c_{N1} = min\{c_{Ni}\}$.
The light neutrino mass is proportional to the charged
lepton mass square,
\begin{equation}
M_\nu=\eta^2 \epsilon^{2c_{N1}} C_L^{-2} M_e^2  ,
\end{equation}
where $ C_L\simeq F_R(c_{E3})$ and $\eta^2 \equiv v_{W}/ (\lambda_L T)$.
The lightest bulk neutrino mass becomes
\begin{equation}
c_{N1} \simeq  \frac{ 10 +\ln(\eta C_L^{-1})}{\ln(T/k)}.
\end{equation}
If $\eta C_L^{-1}\sim 1$,
the bulk neutrino mass is $c_{N1}\simeq
-0.28$, which is quite different from the Dirac neutrino case. While the
charged lepton bulk mass is the same, the Majorana neutrino case contains much
larger bulk masses.
However,
to achieve the MNS matrix (\ref{Un}), $2n \lsim 1.5$ is required
in Majorana neutrino case even for the maximally hierarchical
case ($m_1^\nu=0$). The condition yields  $U_{e3} \sim 0.3 $ which is over the experimental bound
$0.16$. Even with the maximal ambiguity in the approximation, the value is
marginally allowed. It is difficult to match the current experimental data with
the Majorana neutrino in the almost universal Yukawa coupling model.

\section{Conclusions}

In the warped bulk SM, the fermion mixing and mass hierarchy can be induced from
the suppressed zero mode of KK field at the physical boundary.
In the case where the Yukawa couplings are almost universal to all bulk
fermions,
we can
determine the allowed regions of the bulk fermion masses through
the data of  mixings and masses of the SM particles
 with a simple analytic method.

If the Yukawa couplings of all SM fermions
are universal
and if there is no large cancellation in the multiplications
between the different fermion mixing matrices, the current experimental data
almost determine the bulk quark masses.
For the bulk lepton masses, it yield a wide range of solutions.
The existing data cannot narrow down the solution range much.
Still, a few interesting predictions have been found in the lepton sector.
One of them is that only the normal hierarchy is valid neutrino mass hierarchy in this
model.  Another is the fact that it is favorable to consider the
light neutrinos are Dirac fermions. It is because that the seesaw mechanism in
this model generates too large $U_{e3}$.

One of the most notable predictions is on the MNS matrix component
$U_{e3}$ which is predicted  to be $\sim 0.1$. This value
is not so far from the current experimental upper bound and can be tested by
neutrino oscillation experiments in
 near future.
The bulk quarks and lepton masses may be explored at future  high energy
colliders. The third generation of charged fermions
has significantly different features  from others and thus can be a probe for
the  bulk SM.

\acknowledgements
\noindent S.C. was supported
by the Korea Research Foundation Grant funded
by the Korean Government (MOEHRD) No. KRF-2005-070-C00030.
C.S.K. was supported
by the Korea Research Foundation Grant funded
by the Korean Government (MOEHRD) No. R02-2003-000-10050-0.
M.Y. was  supported by the Scientific Grants from the Ministry
of Education, Science, Sports, and Culture of Japan, No.~16081202 and 17340062.

\end{document}